\documentclass[epjmod,nopacs,a4paper,10pt]{svjourmod}
\newcommand{\Lumi}{\mathcal{L}}                            
\newcommand{\lvec}{\langle}                                %
\newcommand{\rvec}{\rangle}                                %
\newcommand{\xF}{x_\mathrm{F}}
\newcommand{\pT}{p_\mathrm{T}}
\newcommand{\pTsq}{p^2_\mathrm{T}}
\newcommand{\xb}{x_\mathrm{b}}

\usepackage{graphicx}
\newcommand{\br}{\hline\noalign{\smallskip}}
\newcommand{\mr}{\noalign{\smallskip}\hline\noalign{\smallskip}}
\newcommand{\bbr}{\noalign{\smallskip}\hline}
\newcommand{\EJP}{{ Eur. J. Phys.} } 
 
\newcommand{\jpg}{{ J. Phys. G: Nucl. Part. Phys.} }   

\newcommand{\NIM}{{ Nucl. Instrum. Methods\/} }
\newcommand{\NP}{{ Nucl. Phys.} }
\newcommand{\PL}{{ Phys. Lett.} }
\newcommand{\PR}{{ Phys. Rev.} }
\newcommand{\PRL}{{ Phys. Rev. Lett.} }

\newcommand{\ZP}{{ Z. Phys.} }

\begin{document}
\hugehead
\title{Measurement of $D^0$, $D^+$, $D_s^+$  and $D^{*+}$ Production 
       in Fixed Target 920~GeV Proton-Nucleus Collisions}

\author{
I.~Abt\inst{24}\and
M.~Adams\inst{11}\and
M.~Agari\inst{14}\and
H.~Albrecht\inst{13}\and
A.~Aleksandrov\inst{30}\and
V.~Amaral\inst{9}\and
A.~Amorim\inst{9}\and
S.~J.~Aplin\inst{13}\and
V.~Aushev\inst{17}\and
Y.~Bagaturia\inst{13\and37}
% ...
}

\mail{marko.staric@ijs.si}

\institute{
$^{1}${\it NIKHEF, 1009 DB Amsterdam, The Netherlands~$^{a}$} \\
$^{2}${\it Department ECM, Faculty of Physics, University of Barcelona, E-08028 Barcelona, Spain~$^{b}$} \\
$^{3}${\it Institute for High Energy Physics, Beijing 100039, P.R. China} \\
$^{4}${\it Institute of Engineering Physics, Tsinghua University, Beijing 100084, P.R. China} \\
$^{5}${\it Institut f\"ur Physik, Humboldt-Universit\"at zu Berlin, D-12489 Berlin, Germany~$^{c,d}$} \\
$^{6}${\it Dipartimento di Fisica dell' Universit\`{a} di Bologna and INFN Sezione di Bologna, I-40126 Bologna, Italy} \\
$^{7}${\it also from Fondazione Giuseppe Occhialini, I-61034 Fossombrone(Pesaro Urbino), Italy} \\
$^{8}${\it Department of Physics, University of Cincinnati, Cincinnati, Ohio 45221, USA~$^{e}$} \\
$^{9}${\it LIP Coimbra, P-3004-516 Coimbra,  Portugal~$^{f}$} \\
$^{10}${\it Niels Bohr Institutet, DK 2100 Copenhagen, Denmark~$^{g}$} \\
$^{11}${\it Institut f\"ur Physik, Universit\"at Dortmund, D-44221 Dortmund, Germany~$^{d}$} \\
$^{12}${\it Joint Institute for Nuclear Research Dubna, 141980 Dubna, Moscow region, Russia} \\
$^{13}${\it DESY, D-22603 Hamburg, Germany} \\
$^{14}${\it Max-Planck-Institut f\"ur Kernphysik, D-69117 Heidelberg, Germany~$^{d}$} \\
$^{15}${\it Physikalisches Institut, Universit\"at Heidelberg, D-69120 Heidelberg, Germany~$^{d}$} \\
$^{16}${\it Department of Physics, University of Houston, Houston, TX 77204, USA~$^{e}$} \\
$^{17}${\it Institute for Nuclear Research, Ukrainian Academy of Science, 03680 Kiev, Ukraine~$^{h}$} \\
$^{18}${\it J.~Stefan Institute, 1001 Ljubljana, Slovenia~$^{i}$} \\
$^{19}${\it University of Ljubljana, 1001 Ljubljana, Slovenia} \\
$^{20}${\it University of California, Los Angeles, CA 90024, USA~$^{j}$} \\
$^{21}${\it Lehrstuhl f\"ur Informatik V, Universit\"at Mannheim, D-68131 Mannheim, Germany} \\
$^{22}${\it University of Maribor, 2000 Maribor, Slovenia} \\
$^{23}${\it Institute of Theoretical and Experimental Physics, 117218 Moscow, Russia~$^{k}$} \\
$^{24}${\it Max-Planck-Institut f\"ur Physik, Werner-Heisenberg-Institut, D-80805 M\"unchen, Germany~$^{d}$} \\
$^{25}${\it Dept. of Physics, University of Oslo, N-0316 Oslo, Norway~$^{l}$} \\
$^{26}${\it Fachbereich Physik, Universit\"at Rostock, D-18051 Rostock, Germany~$^{d}$} \\
$^{27}${\it Fachbereich Physik, Universit\"at Siegen, D-57068 Siegen, Germany~$^{d}$} \\
$^{28}${\it Institute for Nuclear Research, INRNE-BAS, Sofia, Bulgaria} \\
$^{29}${\it Universiteit Utrecht/NIKHEF, 3584 CB Utrecht, The Netherlands~$^{a}$} \\
$^{30}${\it DESY, D-15738 Zeuthen, Germany} \\
$^{31}${\it Physik-Institut, Universit\"at Z\"urich, CH-8057 Z\"urich, Switzerland~$^{m}$} \\
$^{32}${\it visitor from Dipartimento di Energetica dell' Universit\`{a} di Firenze and INFN Sezione di Bologna, Italy} \\
$^{33}${\it visitor from P.N.~Lebedev Physical Institute, 117924 Moscow B-333, Russia} \\
$^{34}${\it visitor from Moscow Physical Engineering Institute, 115409 Moscow, Russia} \\
$^{35}${\it visitor from Moscow State University, 119992 Moscow, Russia} \\
$^{36}${\it visitor from Institute for High Energy Physics, Protvino, Russia} \\
$^{37}${\it visitor from High Energy Physics Institute, 380086 Tbilisi, Georgia} \\
$^\dagger${\it deceased} 
\vspace{5mm}\\
$^{a}$ supported by the Foundation for Fundamental Research on Matter (FOM), 3502 GA Utrecht, The Netherlands \\
$^{b}$ supported by the CICYT contract AEN99-0483 \\
$^{c}$ supported by the German Research Foundation, Graduate College GRK 271/3 \\
$^{d}$ supported by the Bundesministerium f\"ur Bildung und Forschung, FRG, under contract numbers 05-7BU35I, 05-7DO55P, 05-HB1HRA, 05-HB1KHA, 05-HB1PEA, 05-HB1PSA, 05-HB1VHA, 05-HB9HRA, 05-7HD15I, 05-7MP25I, 05-7SI75I \\
$^{e}$ supported by the U.S. Department of Energy (DOE) \\
$^{f}$ supported by the Portuguese Funda\c c\~ao para a Ci\^encia e Tecnologia under the program POCTI \\
$^{g}$ supported by the Danish Natural Science Research Council \\
$^{h}$ supported by the National Academy of Science and the Ministry of Education and Science of Ukraine \\
$^{i}$ supported by the Ministry of Education, Science and Sport of the Republic of Slovenia under contracts number P1-135 and J1-6584-0106 \\
$^{j}$ supported by the U.S. National Science Foundation Grant PHY-9986703 \\
$^{k}$ supported by the Russian Ministry of Education and Science, grant SS-1722.2003.2, and the BMBF via the Max Planck Research Award \\
$^{l}$ supported by the Norwegian Research Council \\
$^{m}$ supported by the Swiss National Science Foundation \\
}


\abstract{
The inclusive production cross sections of the charmed mesons  
$D^0, D^+, D_s^+$ and $D^{*+}$
have been measured in interactions of 920~GeV protons on C, Ti, and W 
targets with the HERA-B detector at the HERA storage ring. Differential 
cross sections as a function of transverse momentum and Feynman's x variable are given for
the central rapidity region and for transverse momenta up to 
$\pT=3.5$~GeV/$c$. The atomic mass number dependence and the leading to non-leading 
particle production asymmetries are presented as well.
} 

\titlerunning{Measurement of $D^0$, $D^+$, $D_s^+$  and $D^{*+}$ Production in 920~GeV $pA$ Collisions}

\maketitle

\section{Introduction}

The cross sections for charm and beauty hadro-production are of considerable 
theoretical interest \cite{lourenco,frixi,nason89}. 
Perturbative QCD is expected to work well for the
large mass top quark production and less well for the lower mass
$b$ and $c$ quarks
\cite{lourenco,frixi,nason89}. 
 At present, several published results of measurements 
of charm production in proton-nucleus collisions \cite{NA16,NA27,E743,E653,E789,E769}
are available. 
They are mainly restricted to beam energies between 200~GeV and 800~GeV and mostly 
have low statistics. Only one of the experiments also provides a 
measurement of the  dependence of the cross section on the atomic mass number. 
More data would help in determining the 
strong interaction parameters as well as in guiding the calculation of 
non-perturbative effects. Another motivation comes from the
prediction that  a prominent manifestation of the quark gluon plasma
at the LHC is a larger ratio of  charmonium
to  open charm cross section compared to, e.g., production in $pA$
collisions at  lower energies~\cite{ccjpsi}. 
The present work provides a new data point at 920~GeV proton
beam energy. 

Collisions of the 920~GeV HERA proton beam in C, Ti and W fixed targets
have been measured with the HERA-B spectrometer. In previous papers
we have reported on the $b \overline{b}$  \cite{bbx}, $\Upsilon$ \cite{ups},
and charmonium  \cite{jpsi,psi-prime} production cross sections,
while the present work deals with production of open charm in the 
inclusive reactions $pA \to D X$. Here 
$D$ represents a $D^0, D^+, D_s^+$ or $D^{*+}$ detected through the 
decay channels: $D^0 \to K^- \pi^+$, $D^+ \to K^- \pi^+ \pi^+$, 
$D_s^+ \to \phi \pi^+ \to K^- K^+ \pi^+$, and $D^{*+} \to D^0 \pi^+ \to K^- \pi^+ \pi^+$. 
Throughout this paper,
        charge-conjugated modes are included unless noted otherwise.

The paper is organized as follows. 
We first briefly describe the apparatus,  the data sample
and the method of analysis. We then present the results and finally  
make a comparison with other measurements and
theoretical expectations.

\section{The detector}

\begin{figure*}[hbt]
\centerline{\includegraphics[width=15cm,clip]{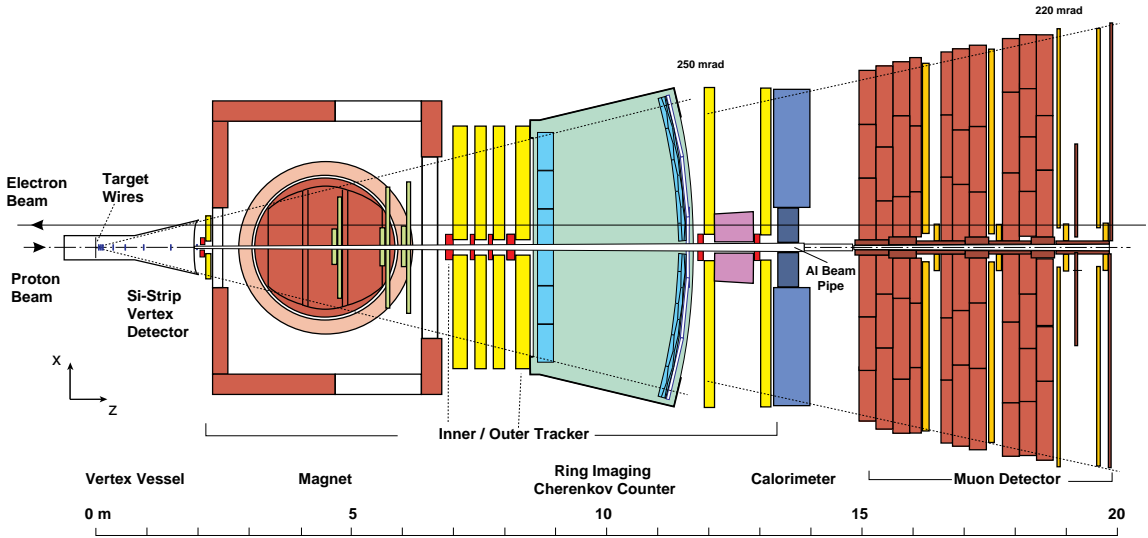}}
\caption[kk]{A top  view of the HERA-B detector.}
\label{hbdet}
\end{figure*}

HERA-B was a fixed target spectrometer (see Fig.~\ref{hbdet})
using the 920~GeV proton beam of
 the HERA $e$-$p$ collider. Interactions occurred on one or more wires (depending on run configuration) which were organized into two groups of four target wires each;     
the groups were separated by 4~cm along the beam, and the  transverse wire dimension 
was  50~$\mu$m-500~$\mu$m~\cite{Target}.
   The wires were positioned in the beam halo,   and their distance to
   the beam core was automatically adjusted to maintain a constant
   interaction rate.
Details of the
various subdetectors have been published \cite{VDS,ITR,OTR,RICH,ECAL,MUON}, 
so only a brief overview of the apparatus is given here.

Tracks originating from proton interactions and decay vertices were measured with a vertex detector
system (VDS) \cite{VDS}. Sixty-four
 silicon strip detectors ($50\times70$~mm$^2$, pitch of 
$\sim$50~$\mu$m) with double-sided readout were arranged 
in eight stations between 7\,cm and 200\,cm downstream of the targets.
The detectors were in Roman pots~\cite{roman-pots} under vacuum and their 
inner edges were 
adjusted to be in the range 10-15\,mm from the beam center. 
With this system, a vertex resolution of $\sigma_z\sim 500~\mu$m 
along the beam direction and $\sigma_{x,y}\sim 50~\mu$m in the transverse plane was achieved.

Particle momenta were measured with a tracking system and a dipole magnet
of 2.13~T$\cdot$m field integral. 
The first tracking chamber was upstream of the magnet, and the remaining 
six chambers
were downstream of the magnet between 7\,m and 13\,m  from the interaction
  region.
 Due to a large variation of particle flux density, the tracking
system was divided into a fine-grained inner tracker (ITR), using microstrip
gas chambers with gas electron multipliers and $\sim$300~$\mu$m pitch \cite{ITR},
 and a coarse-grained outer tracker 
(OTR), using honeycomb drift cells with 5~mm and 10~mm cell diameters \cite{OTR}.
The obtained momentum resolution can be parameterised as 
 $\sigma_p/p = (1.61 + 0.0051 \cdot p~[\mbox{GeV}/c])~\%$~\cite{OTR}, where $p$ is the 
particle momentum.

Particle identification was achieved with three subdetectors: a ring imaging
Cherenkov counter (RICH), an electromagnetic calorimeter (ECAL), and 
muon chambers (MUON). The RICH detector \cite{RICH} 
is a large vessel containing about
100~m$^3$ of C$_4$F$_{10}$ gas at STP, which provided about 2~m of radiation
path. The Cherenkov photons were focused and reflected by two sets of
spherical and planar mirrors onto an upper and a lower photon detector, 
located well outside of the main particle flux.
Each photon detector consisted of about 1100 multi-anode photomultiplier  
tubes. 
The identification efficiency for pions was about 90\% in the momentum range from slightly above the 
pion threshold (2.4~GeV/$c$) up to 70~GeV/$c$, with the 
kaon misidentification probability always below 10\%. The efficiency for kaons
 was above 85\% for momenta between 15~GeV/$c$ and 45~GeV/$c$, with a 
pion misidentification probability of  $\sim$1\%. For particles below the Cherenkov 
threshold of $\sim$10~GeV/$c$, the misidentification of pions as kaons was kept below 10\%.

The ECAL  \cite{ECAL} was a sampling calorimeter of the ``shashlik'' type, with scintillator
plates sandwiched between tungsten (inner region) or lead (outer region)
absorbers. The calorimeter was read out by optical fibers and photomultiplier tubes
with the readout granularity of inner and outer regions
adapted to different particle rates  in order to maintain
acceptable occupancies. The MUON system \cite{MUON}, situated in the most downstream
region, consisted of four large detector stations separated by
concrete and iron absorbers. Each muon detector plane had gas pixel chambers in the
inner region and gas proportional tubes in the outer region.

For the present measurement, the analysis of data was based on the 
 vertex detector, the OTR tracking system and the RICH counter.

\section{Data analysis}

The analysis was performed on data sets with a 
single  target wire made either of carbon, titanium or tungsten. Only 
runs with stable conditions and a minimum bias trigger were considered; 
the resulting sample consisted of 
182 million interactions (Table \ref{data_lumi}). 
The trigger required at least 20 hits in
the RICH detector (compared to an average of 33 for a full ring from a $\beta = 1$
particle \cite{RICH}) 
and had an efficiency $\epsilon_{\rm trigger} \approx 95$\% for inelastic interactions.
The integrated luminosity $\cal{L}$ was determined \cite{lumi} from the number of 
measured inelastic
interactions $N_{\rm inel}$ and the total inelastic cross section $\sigma_{\rm inel}$,
  using the expression 
$\mathcal{L} = N_{\rm inel}/(\epsilon_{\rm trigger}\sigma_{\rm inel})$. 
The data were
recorded at a moderate interaction rate of about 1.5~MHz which corresponds
to 0.17 interactions per filled bunch crossing. Therefore only about 10\% of 
triggered
events contain more than one interaction. The data acquisition rate was
about 1~kHz, and the bulk of the data  was recorded within a two-week  running period.

\begin{table}
  \caption{Summary of the data statistics and the integrated luminosities
    of the present study. 
}
   \begin{center}
    \begin{tabular}{@{}llll}
      \br
      Target  & A & events [$\times 10^6$] & $\mathcal{L}$ [$\mu$b$^{-1}$] \\
      \mr
      C  &   12.01  & 89.3 & 375 \\
      Ti &   47.88  & 24.7 & 31   \\
      W  &  183.84  & 67.6 & 36  \\
      \bbr
    \end{tabular}
   \end{center}
  \label{data_lumi}
\end{table}

At the HERA-B energy, the charm production cross section is  more than
two orders
of magnitude smaller than the inelastic cross section. Taking into account  
the relatively small branching ratios for the $D$ meson decay modes into two  or three charged 
particles, 
one expects sizeable backgrounds. 
Particle identification alone is not sufficient to extract signal events. 
However, the large boost of the
center-of-mass system of HERA-B ($\gamma=22$), causing $D$ mesons to decay 
several 
millimeters from the target, combined with  good vertex resolution ($\approx$0.5\,mm longitudinally) allows us
to distinguish $D$ meson decay products from 
particles originating at the primary interaction point.
The data selection thus requires a detached secondary vertex formed by tracks
not coming from the primary interaction point as well as the identification 
of kaons and pions. The selection criteria are summarized in Table \ref{AnalCuts.tab} and discussed in 
some detail below. 

\subsection{Data selection}

In addition to a detached vertex for the ground state $D$ mesons, 
at least one reconstructed primary vertex was required in each selected event. 
Primary vertices were determined from all track segments reconstructed in the VDS.
Since the proton interaction point must be inside the target wire,
 the primary vertex coordinate transverse to the wire
direction was replaced with the known target wire position. 

For the  tracks corresponding to the  decay products of $D$ mesons,
mild requirements were applied on the number of hits in the vertex detector and in 
the main tracking system, as well as on the track fit quality. 
Selection criteria used 
to identify the final state kaons and pions were based on information from the RICH counter.
While a rather strict cut  on the kaon likelihood\footnote{Likelihoods for the electron, muon, pion, 
kaon, proton and background hypotheses are normalized such that their sum is equal to one: 
$L_e+L_{\mu}+L_{\pi}+L_K+L_p+L_{\rm bkg}=1$.}  was required for the kaon candidates
($L_K > 0.5$ for the $D^0$ and $D^+$ selection, and $L_K > 0.33$ for $D_s^+$  candidates), for pions
only a mild cut was applied on the sum of RICH likelihoods 
for electrons, muons and pions,  $L_e+L_{\mu}+L_{\pi}>0.05$. 
No particle identification requirement was imposed 
for pions from $D^{*+} \to D^0 \pi^+ $ decays, which tend to have low momenta and
are thus denoted as $\pi_{\rm slow}$.

The tracks were combined to form $D^0, D^+$ and $D_s^+$ candidates. Candidates 
with an invariant mass in the interval of 
$\pm$500~MeV/$c^2$ around the $D$ meson nominal mass were retained for further 
analysis. 
For the $D_s^+$ candidates, the 
invariant mass of the $K^+K^-$ pairs was required to be in the interval of  $\pm$10~MeV/$c^2$ 
around the $\phi$ nominal mass; the absolute value of the cosine of the angle $\theta_\phi$
between the $K^+$ and $\pi^+$ in the rest frame of the $\phi$ was restricted to the 
values above 0.5, exploiting the vector nature of the intermediate state $\phi$.  
The $D^{*+}$ candidates were 
reconstructed from $D^0$ candidates with invariant mass within  $\pm$75~MeV/$c^2$ 
($3.5\sigma$ in resolution)  of the 
$D^0$ mass and slow pion candidates, after a vertex fit to the $D^0$;
 an additional cut was applied on the product of transverse momenta
 of the $D^0$ daughter tracks and of the $D^0$ momentum, $p(D^0)\pT(K)\pT(\pi)$.

In addition to the criteria described above, the analysis was restricted  to the 
region of phase space with high acceptance, $  -0.15 < \xF < 0.05$, where  
$\xF$ is the Feynman $x$ variable. 
The daughter tracks of a $D$ meson candidate  were fitted to a common vertex. Only 
combinations  with a vertex probability greater than 0.1\% and with a
secondary vertex displaced by more than  4 standard deviations downstream 
of the wire were accepted.

The $D$ meson candidate was then associated with the primary vertex. In the case 
of events 
with  multiple reconstructed primary vertices, the vertex with the smallest
impact parameter significance (i.e., the measured value 
 divided by its estimated error given by the covariance matrix)
with respect to the track of the $D$ meson candidate was chosen. 
To avoid a possible bias in the primary vertex position due to tracks from the 
$D$ meson candidate, 
the primary vertex was re-fitted without the $D$ daughter tracks.

The final set of criteria was based on the primary and secondary vertices. 
The main source of background, which is due to combinations of particles emerging 
from 
the primary interaction point, was reduced by the following
requirements: (1) the secondary vertex should be detached, (2) the tracks forming 
the
secondary vertex should not come from the primary interaction point and (3) the
$D$ meson candidate should originate from the primary interaction point. To fulfill these 
criteria, cuts were applied on the following variables:
\begin{itemize}
\item $d(D)$ the significance of the distance between the secondary vertex and  
the associated primary vertex,
\item $b(\pi), b(K)$ the significance of the impact parameter of a pion or kaon  
with respect to the  primary vertex (in case more than one primary vertex was found, that 
closest to the particle is chosen),
\item $b(D)$ the  significance of the impact parameter of a $D$ meson candidate 
with respect to the associated primary vertex. 
\end{itemize}

We found that, for the three-body decays of $D^+$ and $D_s^+$, a more effective
cut than a selection based on single impact parameter significances of
daughter tracks is a cut on their product $b(K)b(\pi_1)b(\pi_2)$ and 
$b(K_1)b(K_2)b(\pi)$, respectively. 
The background level is further reduced with criteria of the form
$\sqrt[3]{b(K)b(\pi_1)b(\pi_2)} > 4(t-t_0)$, for $D^+$, and 
 $\sqrt{b(\phi)b(\pi)} > 0.75(t-t_0)$, for $D_s^+$. 
Here the proper lifetime $t$ is in units of the $D$ mean lifetime, and the  offset $t_0$ 
is determined in an optimization.

For each decay mode the optimal cuts were determined by maximizing the signal significance
 $S/\sqrt{S+B}$ in a $\pm 3\sigma$ window centered at the $D$ meson nominal mass (signal window).
The signal $S$ was taken from Monte Carlo simulation 
and was scaled to the luminosity of real data by using an estimation
for the production cross sections from fits to the published $D$ meson
cross sections \cite{lourenco}. For the $D_s^+$ the cross section was 
assumed to amount to 20\% of the sum of the $D^0$ and $D^+$ cross sections \cite{frixi}. 

In the case of ground state $D$ mesons, the number of background events $B$ 
was estimated from the data sidebands. 
For the $D_s^+$, the mass region of $\pm50$~MeV/$c^2$ around the nominal mass of the $D^+$, where a 
contribution of the decay $D^+ \to \phi \pi^+$ is expected, was excluded from the lower sideband.
To reduce the sensitivity to statistical fluctuations, 
the sidebands were chosen to be larger than the signal window.
In the case of the $D^{*+}$, the wrong sign combinations from real data were used
to estimate the background in the signal window. As is usually done for this 
decay,
the signal was reconstructed
via the mass difference $q=m(K,\pi,\pi_{\rm slow})-m(K,\pi)-m_\pi$ 
rather than using the invariant mass of the $K,\pi,\pi_{\rm slow}$ combinations.

\begin{table*}[htb]
  \caption{Selection criteria;  $d$ and $b$ denote decay distance and 
impact parameter significances,
respectively, and $L_h$ are likelihoods for a hypothesis $h$.}
    \begin{tabular}{@{}ll}
      \br
      $D^0 \to K^-\pi^+$ \phantom{xxxxxxxxxxxxxxxxxxxxxxxxxxxxxxxxxxx} & $D^+ \to K^-\pi^+\pi^+$   \\
      \mr
      \multicolumn{2}{l}{chosen a priori } \\
      \mr
      $L_K(K)>0.5$    & $L_K(K)>0.5$ \\
      $L_e(\pi)+L_{\mu}(\pi)+L_\pi(\pi)>0.05$ & $L_e(\pi)+L_{\mu}(\pi)+L_\pi(\pi)>0.05$  \\
      \mr
      \multicolumn{2}{l}{optimized using background data and signal MC} \\
      \mr
      $d(D^0)>6.1$      & $b(D^+)<2.6$ \\
      $b(D^0)<2.4$      & $b(K)b(\pi_1)b(\pi_2)>106$ \\
      $b(K)>3.4$    & $\sqrt[3]{b(K)b(\pi_1)b(\pi_2)} > 4(t-t_0),~t_0=2.48$ \\
      $b(\pi)>3.7$  & \\
      \br
      \multicolumn{2}{l}{} \\
      \br
      $D_s^+ \to \phi \pi^+ \to K^-K^+\pi^+$ & $D^{*+} \to D^0\pi^+ \to K^-\pi^+\pi^+$ \\
      \mr
      \multicolumn{2}{l}{chosen a priori } \\
      \mr
      $L_K(K)>0.33$            & $L_K(K)>0.5$ \\
      $L_e(\pi)+L_{\mu}(\pi)+L_\pi(\pi)>0.05$          & $L_e(\pi)+L_{\mu}(\pi)+L_\pi(\pi)>0.05$ \\
      $|m(K^+K^-)-m(\phi)|<10$~MeV/$c^2$  & $|m(K\pi)-m(D^0)|<75$~MeV/$c^2$ \\
      $|\cos (\theta_\phi)|>0.5$ & $d(D^0)>4$ \\
      \mr
      \multicolumn{2}{l}{optimized using background data and signal MC} \\
      \mr
      $d(D_s^+)>5.3$                            & $b(D^0)<2.4$ \\
      $b(D_s^+)<2.11$                           & $b(K)>2.1$ \\
      $b(K^-)b(K^+)b(\pi)>28.7$           & $b(\pi)>1.7$ \\
      $\sqrt{b(\phi)b(\pi)} > 0.75(t-t_0),~t_0=1.0$    & $p(D^0)\pT(K)\pT(\pi)>17.7({\rm GeV}/c)^3$ \\
      \br
    \end{tabular}
  \label{AnalCuts.tab}
\end{table*}

After applying the selection criteria, which are summarized in Table \ref{AnalCuts.tab}, 
the remaining data were scanned for events with more 
than one $D$ meson candidate (in 0.5\%, 8\%, 0\% and 20\% of events 
for the $D^0$, $D^+$, $D^+_s$ and $D^{*+}$  candidates, respectively).
In case of the $D^0$, $D^+$ and $D^+_s$ candidates, the combinations with the 
largest decay distance significance $d(D)$ were kept. For the $D^{*+}$ analysis, first the 
candidates whose intermediate $D^0$ had the largest decay distance 
significance were selected. If multiple candidates remained (i.e., due to multiple  $\pi_{\rm slow}$ candidates),
the $D^{*+}$  candidate with the highest vertex probability was kept.

\subsection{Signal yields}

\begin{figure}[ht]
  \centerline{\includegraphics[width=8cm]{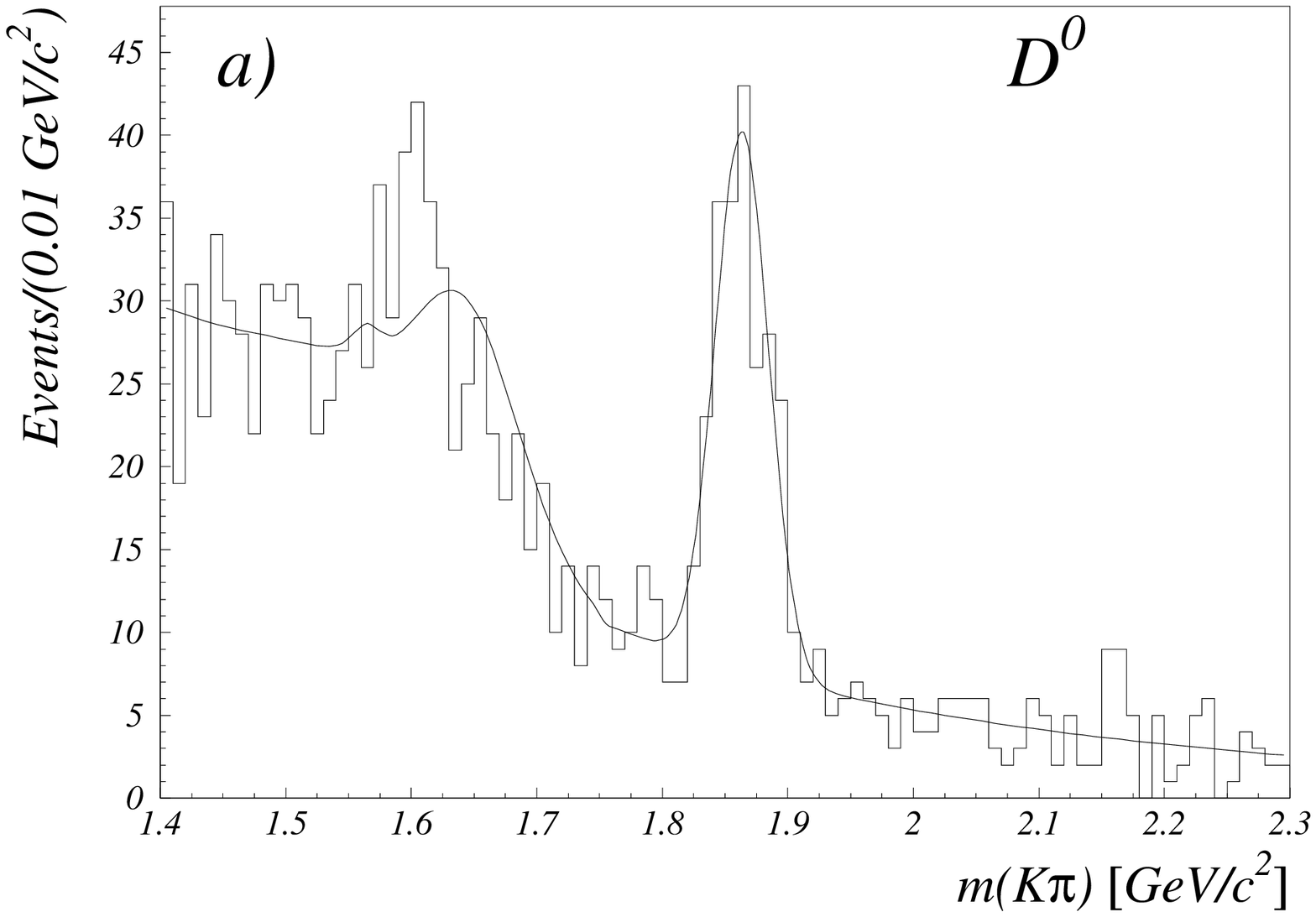}}
  \centerline{\includegraphics[width=8cm]{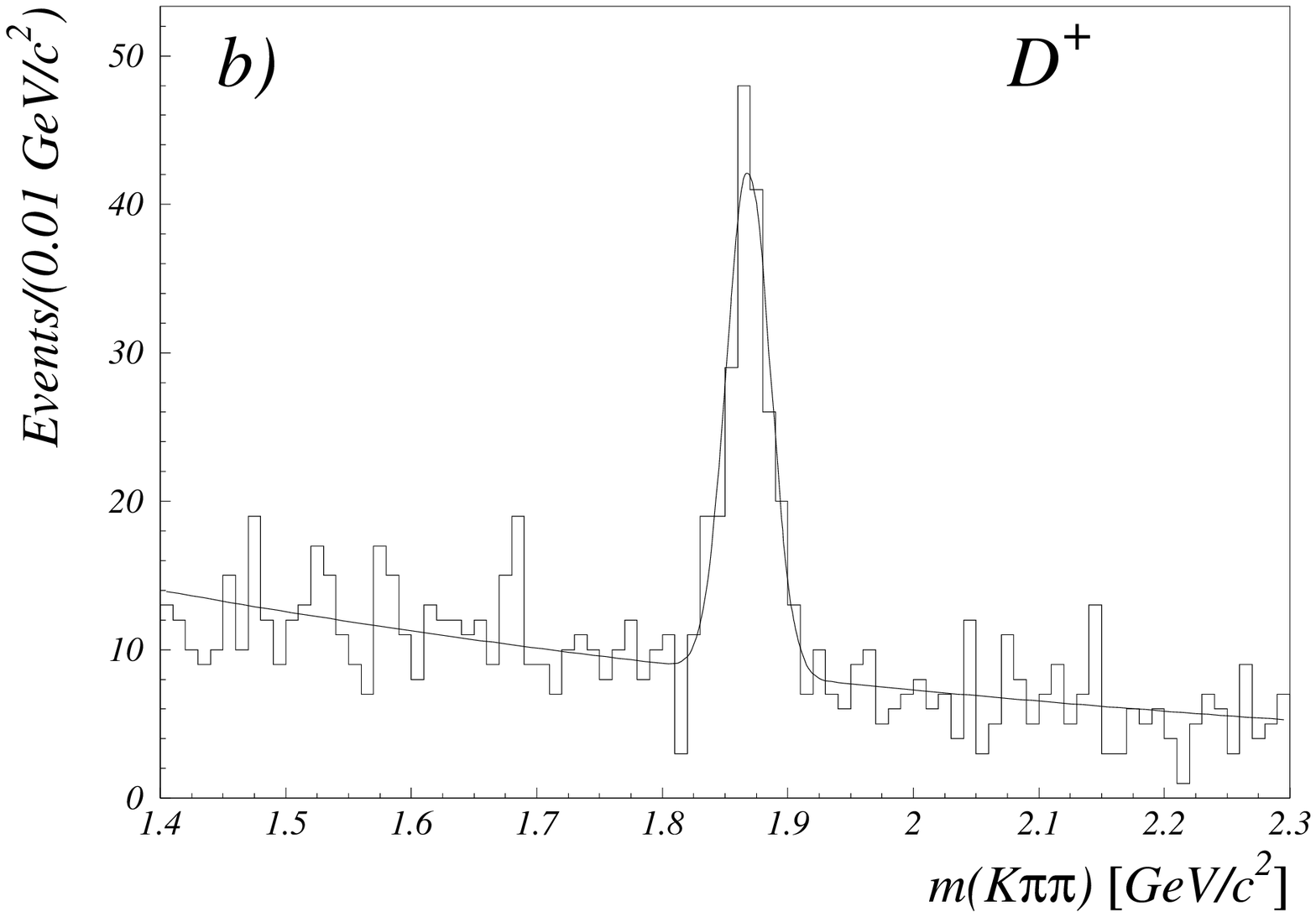}}
  \caption{Invariant mass distributions for $K^-\pi^+$ (a) and 
$K^-\pi^+\pi^+$ (b) combinations. The curves show  results of  
 maximum likelihood fits to the data.}
  \label{mass_d0dp.eps}
\end{figure}

\begin{figure}[ht]
  \centerline{\includegraphics[width=8cm]{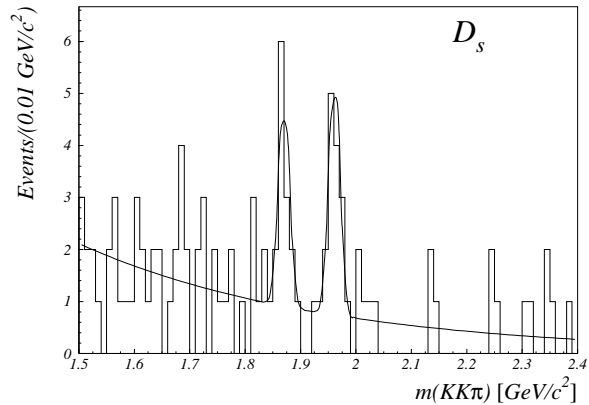}}
  \caption{Invariant mass distributions for $\phi\pi^+ \to (K^-
K^+)\pi^+$ combinations. 
    Besides the $D_s^+$ peak at 1.96~GeV/$c^2$, a $D^+$ peak at 1.87~GeV/$c^2$ is also visible.
    This peak corresponds to the Cabibbo suppressed decay $D^+ \to \phi \pi^+$. 
    The curve shows  the result of a  maximum likelihood fit to the data.}
  \label{mass_ds.eps}
\end{figure}

\begin{figure}[ht]
  \centerline{\includegraphics[width=8cm]{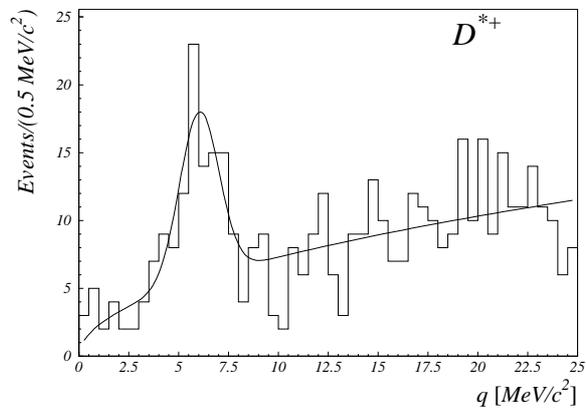}}
  \caption{Invariant mass difference $q=m(K,\pi,\pi_{\rm slow})-m(K,\pi)-m_\pi$
  for $K^-\pi^+\pi^+$ combinations. The curve shows  the result of a 
  maximum likelihood fit to the data.}  
  \label{mass_dstar.eps}
\end{figure}

The invariant mass distributions for $D$ meson candidates after 
applying the selection criteria discussed above are shown in 
Figs.~\ref{mass_d0dp.eps}-\ref{mass_dstar.eps}.

The signal yields are extracted from  
the histograms by a maximum likelihood fit assuming Poisson statistics in individual
bins. A Gaussian function is used for the signal,
 while the background description depends on the type of the $D$ meson. 
For the $D^+$ and $D_s^+$, the background is fitted by an exponential function. 
The
background for $D^0$ candidates is more complex and consists of a 
combinatorial part, fitted by an exponential, and a contribution  from partially 
reconstructed charm decays. This latter background is visible in the mass range below 
the $D^0$ peak. 
The shape of this background is taken from  Monte Carlo simulation of $c\bar c$ 
events to which the same selection criteria are applied as for the data. 

In the $D_s^+$ invariant mass distribution the Cabibbo suppressed decay of
$D^+ \to \phi\pi^+$ is also seen (the peak to the left of the  $D_s^+$ peak). 
This peak is included in the fit function as an additional Gaussian of 
the same width as the signal Gaussian, with its normalization as an additional 
free parameter, and  its 
mean    fixed to that  extracted from the $D^+ \to K^-\pi^+\pi^+$ 
invariant mass distribution.
The background for the $D^{*+}$ candidates is parameterised as
$a(q^{1/2} + bq^{3/2})$, with  $a$ and $b$ as free parameters.

The fitted peak positions are within one standard deviation of the 
corresponding world average 
values \cite{PDG}, with the exception of $D_s^+$, which deviates by less than
two standard deviations. 
The widths of the signal peaks are about 30\% larger than the corresponding
Monte Carlo values. 

The numbers of reconstructed $D$ mesons are summarized in Table \ref{num_d_data.tab}. 
In total, 175 $D^0$, 131 $D^+$, 11 $D_s^+$ and 61 $D^{*+}$ 
decays are found.
The yields for various subsamples (particle, anti-particle, individual
target material) are obtained by fitting with the mean and the width (r.m.s.) of the signal Gaussian function
fixed to the values obtained from the fit to the full sample.

\begin{table}
  \caption{Number of reconstructed $D$ mesons.}
   \begin{center}
    \begin{tabular}{lcccc}
      \br
      sample & $D^0$ & $D^+$ & $D_s^+$ & $D^{*+}$ \\
      \mr
total         & 174.8$\pm$16.8  & 130.5$\pm$14.7  &11.4$\pm$4.0  &  61.3$\pm$13.0\\
particle      &75.9$\pm$10.9    &  54.5$\pm$9.3  & 4.9$\pm$2.6  &  21.0$\pm$6.6\\
anti-particle &99.0$\pm$11.9    &  75.8$\pm$10.5  & 6.7$\pm$2.8  &  40.6$\pm$8.3\\
C             &66.1$\pm$9.6     &  43.1$\pm$7.7 & 4.2$\pm$2.2  &  26.6$\pm$6.4\\
W             &92.3$\pm$11.7    &  72.4$\pm$10.6  & 6.7$\pm$3.0  &  24.8$\pm$7.5 \\
Ti            &17.4$\pm$5.7     &  14.9$\pm$5.0  & 0.4$\pm$1.0  &   9.6$\pm$4.0\\
      \bbr
    \end{tabular}
   \end{center}
  \label{num_d_data.tab}
\end{table}

The $D$ meson proper time
distributions provide a check for signal consistency. The acceptance-correc\-ted distributions are 
found to be consistent with the expected
exponential decay. To determine the lifetimes,  a simultaneous likelihood fit of events in the
signal window and in the sidebands is used.  The results,
$c\tau= (302 \pm  33)$~$\mu$m for $D^+$,   $(120 \pm  13)$~$\mu$m 
for  $D^0$, and $(165 \pm  52)$~$\mu$m for  $D_s^+$,   are in good
agreement with the world average values \cite{PDG}. 

As a further consistency check of the  $D_s^+$ and $D^{*+}$ signals, 
the  intermediate states 
$\phi$ and $D^0$ are checked. They should be visible in the 
corresponding invariant mass distributions, when the signal region in the 
initial
state invariant mass distribution is selected, and the fit to the intermediate 
state
invariant mass distribution should give an event yield consistent with the
number of events in the initial state peak.
The yields extracted in this way are in good agreement with the values given in  
 Table \ref{num_d_data.tab}.
For the  $D_s^+$ decays, the distribution of the  cosine of the angle 
between $K^+$ and $\pi$ in the rest frame of the $\phi$ is found to be 
consistent with the expectation given by
 the vector nature of the intermediate state $\phi$.  
The  expected $D^+ \to \phi\pi^+$  signal yield  can be estimated from the number 
of reconstructed decays from our measured $D^+$ cross section in the 
$D^+ \to K^-\pi^+\pi^+$ decay channel. The fitted  number of events, 
9.8$\pm$3.8,  is in reasonable agreement ($1.4\sigma$ higher) with the estimated number, 
(4.2$\pm$1.2).

\section{Efficiency determination}

A Monte Carlo simulation is used to determine the signal reconstruction 
efficiencies. The Monte Carlo samples for $pA \to D X$ are generated in two steps. 
First, a $c\bar c$ pair is generated with Pythia 5.7 \cite{pythia} 
such that a 
particular $D$ meson is always produced. The generated events are re-weighted 
to make the resulting cross sections conform to the parameterisations 
 \begin{eqnarray} 
   \frac{d\sigma}{d\pTsq}&\propto &\left[1+\left(\frac{\sqrt{\pi}~\Gamma(\beta-\frac{3}{2})~\pT}
     {2~\Gamma(\beta-1)~\lvec \pT \rvec}\right)^2\right]^{-\beta}
   \label{ptform.eq}
 \end{eqnarray}
and
\begin{eqnarray}
  \frac{d\sigma}{d\xF} = \left \{ \begin{array}{lcl}
    A\exp{\bigl(-\frac{\xF^2}{2\sigma_g^2}\bigr)} &,& |\xF|<\xb, \\
    A'\bigl(1-|\xF|\bigr)^n &,& |\xF| \ge \xb, 
    \end{array} \right.
  \label{xf-extended.eq}
\end{eqnarray}
with  $\sigma_g=\sqrt{\frac{\xb(1-\xb)}{n}}$ and 
 $ \ln\frac{A}{A'}=n[\frac{\xb}{2(1-\xb)}+\ln(1-\xb)]$ \cite{E791}.
The average transverse momentum of $\lvec \pT \rvec=1.04\pm0.04$~GeV/$c$ and 
 the value of the exponent $\beta=7.0\pm4.3$ 
are  taken from the present analysis (Sec.~\ref{differential.sec}).
The value $n = 7.7 \pm 1.4$ is taken from the average of  E743 and E653 \cite{E743,E653},  
and $\xb$ is assumed to be $0.062\pm0.013$ as measured by the E791 experiment \cite{E791}. 
The influence of the parameter uncertainties and other possible parameterisations 
of Eqs.~\ref{ptform.eq} and \ref{xf-extended.eq} are taken into account as systematic errors. 
After the generation of the $D$ mesons,  the remaining energy is input to 
the Fritiof 7.02 \cite{fritiof} program package 
which generates the underlying event taking into account 
further interactions inside the nucleus.

The detector response is simulated with the Geant 3.21 package \cite{geant}. Realistic 
detector efficiencies, readout noise and dead channels are taken into account. 
The simulated events are processed by the same reconstruction codes used for 
the data. 

\section{Results}
\label{results.sec}

\subsection{Total cross sections}
\label{total.sec}

The visible cross section per nucleus, i.e., the cross section measured in our
visible range of $-0.15<\xF<0.05$, is given by
\begin{eqnarray}
  \Delta \sigma_{{\rm pA},\,i}=\frac{N_i}{{\rm Br} \cdot \epsilon_i \cdot \Lumi_i},
  \label{Dxsec_pA.eq}
\end{eqnarray}
where $N_i$ is the number of reconstructed $D$ mesons for a
particular target $i$, $\epsilon_i$ and $\Lumi_i$ are the corresponding
efficiency and integrated luminosity, 
and ${\rm Br}$ is the world average branching ratio for a specific decay channel
\cite{PDG}.
The cross section for $D$ meson production on a nuclear target of atomic mass number 
$A$
is parameterised as
\begin{eqnarray}
  \sigma_{\rm pA}=\sigma_{\rm pN} \cdot A^\alpha.
  \label{A-dep.eq}
\end{eqnarray}
To combine data recorded with different target materials,  
the production cross sections per nucleon $\Delta \sigma_{\rm pN}$ 
are extracted in the following way.
From Eq.~\ref{Dxsec_pA.eq} and Eq.~\ref{A-dep.eq},  the 
$D$ meson yield of the target $i$ is  derived,
\begin{eqnarray}
  N_i = {\rm Br} \cdot \epsilon_i \cdot \Lumi_i\cdot \Delta \sigma_{\rm pN} \cdot 
A_i^\alpha.
  \label{Ni.eq}
\end{eqnarray}
By summing Eq.~\ref{Ni.eq} over all targets and solving it for the production 
cross section we get
\begin{eqnarray}
  \Delta \sigma_{\rm pN}=\frac{N}{{\rm Br} \cdot \sum_i{\epsilon_i \Lumi_i A_i^\alpha}},
  \label{Dxsec_pN.eq}
\end{eqnarray}
where $N=\sum_i{N_i}$ is the measured $D$ meson yield of the total data sample.  
The sum in the denominator of Eq.~\ref{Dxsec_pN.eq} can be rewritten 
by introducing the average efficiency $\epsilon$, defined by the weighted sum
\begin{eqnarray}
  \epsilon=\sum_i w_i\epsilon_i~~,\qquad
  w_i=\frac{A_i^\alpha\Lumi_i}{\sum_k A_k^\alpha \Lumi_k}.
  \label{av_effi.eq}
\end{eqnarray}
Then the expression \ref{Dxsec_pN.eq} reads
\begin{eqnarray}
  \Delta \sigma_{\rm pN}=\frac{N}{{\rm Br} \cdot \epsilon \cdot \sum_i{\Lumi_i A_i^\alpha}}.
  \label{Dxsec.eq}
\end{eqnarray}
Since there is no experimental indication for nuclear effects in open charm production, 
 a linear $A$ dependence of the cross sections is assumed, and $\alpha$ is set 
to one in Eqs.~\ref{av_effi.eq} and \ref{Dxsec.eq}. This assumption is also 
in agreement with our measurements as discussed in Section~\ref{adep.sec}. 

The total systematic uncertainties are, according to Eq.~\ref{Dxsec.eq}, composed
of contributions from uncertainty in the signal yields associated with the 
fitting procedure (listed as 'event fitting'), branching fractions, integrated luminosity 
and 
reconstruction efficiency. The uncertainty of the reconstruction
efficiency can be further divided into contributions from Monte Carlo statistics, track
reconstruction efficiency, particle identification efficiency, selection criteria
and the contribution from the re-weighting of kinematical distributions based on $\pT$ and $\xF$. 
The individual contributions are summarized in Tables~\ref{syssum.tab}-\ref{syseff.tab}. 
The uncertainty due to selection criteria is determined in two ways:
by varying the cut values in the selection criteria discussed above, and by performing a 
second analysis with an independent set of selection criteria. 
The change in the resulting cross section is taken as
         the corresponding systematic uncertainty.
The systematic errors of track reconstruction and RICH particle
      identification (1.5\% and 2.0\% per track correspondingly) are
      estimated using decays of $K^0_S$ and $\phi$ as sources of pions
      and kaons.
Note that no systematic error is assigned to the assumption  $\alpha=1$ in order to 
be compatible with previous experiments. Using Eq.~\ref{Dxsec_pN.eq} and data in 
Tables \ref{data_lumi} and \ref{xsec.tab}, $\Delta \sigma_{\rm pN}$  
for $\alpha \neq 1$ can be extracted. 

\begin{table}
  \caption{Summary of systematic uncertainties of visible cross sections}
   \begin{center}
    \begin{tabular}{@{}lllll}
      \br
      Source   & $D^0$ & $D^+$ & $D_s^+$ & $D^{*+}$ \\
      \mr
      Event fitting        & 3.4\% & 2.6\% & 6.0\% & 9.7\% \\
      Branching fractions  & 1.8\% & 3.6\% &  13.0\% & 1.9\% \\
      Luminosity           & 3.7\% & 3.7\% & 3.7\% & 3.7\% \\
      Reconstruction efficiency           & 8.4\% &  10.3\% &  12.9\% & 9.7\% \\
      \mr
      Total                &10\% & 12\% & 20\% & 14\% \\
      \bbr
    \end{tabular}
   \end{center}
  \label{syssum.tab}
\end{table}

\begin{table}
  \caption{Summary of systematic uncertainties of reconstruction efficiency.}
   \begin{center}
    \begin{tabular}{@{}lllll}
      \br
      Source   & $D^0$ & $D^+$ & $D_s^+$ & $D^{*+}$ \\
      \mr
      Monte Carlo statistics  & 1.2\% & 1.3\% & 3.7\% & 1.1\% \\
      Track reconstruction    & 3.0\% & 4.5\% & 4.5\% & 4.5\% \\
      Particle identification & 4.0\% & 6.0\% & 6.0\% & 4.0\% \\
      Selection criteria      & 6.0\% & 6.0\% & 6.0\% & 6.0\% \\
      Re-weighting            & 2.9\% & 3.6\% & 7.7\% & 4.6\% \\
      \mr
      Total                   & 8.4\% &10.3\% & 12.9\% & 9.7\% \\
    \bbr
    \end{tabular}
   \end{center}
  \label{syseff.tab}
\end{table}

The resulting cross sections in the visible range, $\Delta \sigma_{\rm pN}$ and $\Delta \sigma_{\rm pA}$,
 are summarized in Tables~\ref{xsec.tab}-\ref{xsec-mat.tab}.
In order to extrapolate the measurements to the full phase space, 
\begin{eqnarray}
  \sigma_{\rm pN}=\frac{\Delta \sigma_{\rm pN}}{f_{\rm vis}},
  \label{fullxsec.eq}
\end{eqnarray}
the fraction $f_{\rm vis}$ of $D$ mesons in the visible range, defined 
by $-0.15<\xF<0.05$,  is determined in the following way. 
With the value of the exponent $n=7.7\pm1.4$  as measured by 
the E743 and E653 experiments \cite{E743,E653},  $f_{\rm vis}$ is calculated
 by using Eq.~\ref{xf-extended.eq}. The values 
corresponding to the choice $x_{\rm b}=0.062$~\cite{E791} and $x_{\rm b}=0$ are 0.542$\pm$0.048 and
 0.558$\pm$0.051, respectively. The difference is small compared to the uncertainty
due to the error of the parameter $n$. For the extrapolation the average  
of both numbers, $ f_{\rm vis} = 0.55\pm0.05$, is used. The resulting cross sections extrapolated
 to the full  phase space are listed in Table \ref{xsec.tab}.

 \begin{table}[b]
   \caption{Cross sections in the visible range ($-0.15<\xF<0.05$) and extrapolated to 
      the full phase space. The first error is statistical and 
the second systematic. In the second column, the systematic error due to the 
extrapolation uncertainty is quoted separately. }
   \begin{center}
     \begin{tabular}{@{}lll}
       \br
       &             $\Delta \sigma_{\rm pN}[\mu{\rm b}]$ (visible range) & $\sigma_{\rm pN}[\mu{\rm b}]$ (full $\xF$ range)  \\
       \mr
     $D^0$           & 26.8$\pm$2.6$\pm$2.7  & 48.7$\pm$4.7$\pm$4.9$\pm$4.4 \\
     $D^+$           & 11.1$\pm$1.2$\pm$1.3  & 20.2$\pm$2.2$\pm$2.4$\pm$1.8 \\
     $D_s^+$         & 10.2$\pm$3.5$\pm$2.0  & 18.5$\pm$6.4$\pm$3.7$\pm$1.7 \\
     $D^{*+}$        & 11.9$\pm$2.6$\pm$1.7  & 21.6$\pm$4.7$\pm$3.0$\pm$2.0 \\
       \bbr
     \end{tabular}
   \end{center}
   \label{xsec.tab}
 \end{table}


\begin{table}[b]
  \caption{Cross sections for particle and anti-particle production 
in the visible range ($-0.15<\xF<0.05$). The first error is statistical and 
the second systematic.}
   \begin{center}
    \begin{tabular}{@{}lrr}
      \br
      \multicolumn{3}{c}{$\Delta \sigma_{\rm pN}[\mu{\rm b}]$} \\
      \mr
      &             particles~~~~          & anti-particles \\
      \mr
      $D^0$        &12.0$\pm$1.7$\pm$1.2   & 14.8$\pm$1.7$\pm$1.5 \\
      $D^+$        & 4.8$\pm$0.8$\pm$0.6   & 6.3$\pm$0.9$\pm$0.8 \\
      $D_s^+$      & 4.1$\pm$2.3$\pm$0.8   & 6.3$\pm$2.6$\pm$1.3 \\
      $D^{*+}$     & 4.5$\pm$1.4$\pm$0.6   & 7.2$\pm$1.5$\pm$1.0 \\
      \bbr
    \end{tabular}
   \end{center}
  \label{xsec-part.tab}
\end{table}

\begin{table}
  \caption{Visible cross sections per nucleus. Numbers in parentheses in 
  the last row are for the subsamples of $D^{*+}$ with $D^0$ daughters not 
  common to the $D^0$ samples of the first row. The first error is statistical and 
the second systematic.}
   \begin{center}
    \begin{tabular}{@{}llll}
      \br
      \multicolumn{4}{c}{$\Delta \sigma_{\rm pA}$ [mb]} \\
      \mr
      data sample & ~~~~~~~~~~C & ~~~~~~~~~~Ti & ~~~~~~~~~~W   \\
      \mr
 $D^0$     &  \phantom{(}0.36$\pm$ 0.05$\pm$ 0.04 &  \phantom{(}1.01$\pm$ 0.33$\pm$ 0.10 &  \phantom{(}4.91$\pm$ 0.62$\pm$ 0.54  \\
 $D^+$     &  \phantom{(}0.12$\pm$ 0.02$\pm$ 0.02 &  \phantom{(}0.48$\pm$ 0.16$\pm$ 0.06 & \phantom{(}2.17$\pm$ 0.32$\pm$ 0.28  \\
 $D^+_s$   &  \phantom{(}0.12$\pm$ 0.07$\pm$ 0.02 &  \phantom{(}0.15$\pm$ 0.35$\pm$ 0.03 &  \phantom{(}2.18$\pm$ 0.99$\pm$ 0.46  \\
 $D^{*+}$  &  \phantom{(}0.17$\pm$ 0.05$\pm$ 0.03 &  \phantom{(}0.71$\pm$ 0.30$\pm$ 0.10  &  \phantom{(}1.71$\pm$ 0.51$\pm$ 0.26  \\
           &            (0.21$\pm$ 0.06$\pm$ 0.03)&            (0.95$\pm$ 0.50$\pm$ 0.14) &            (2.11$\pm$ 0.84$\pm$ 0.34) \\
      \bbr
    \end{tabular}
   \end{center}
  \label{xsec-mat.tab}
\end{table}

\begin{figure*}[t]
  \centerline{\includegraphics[width=15cm]{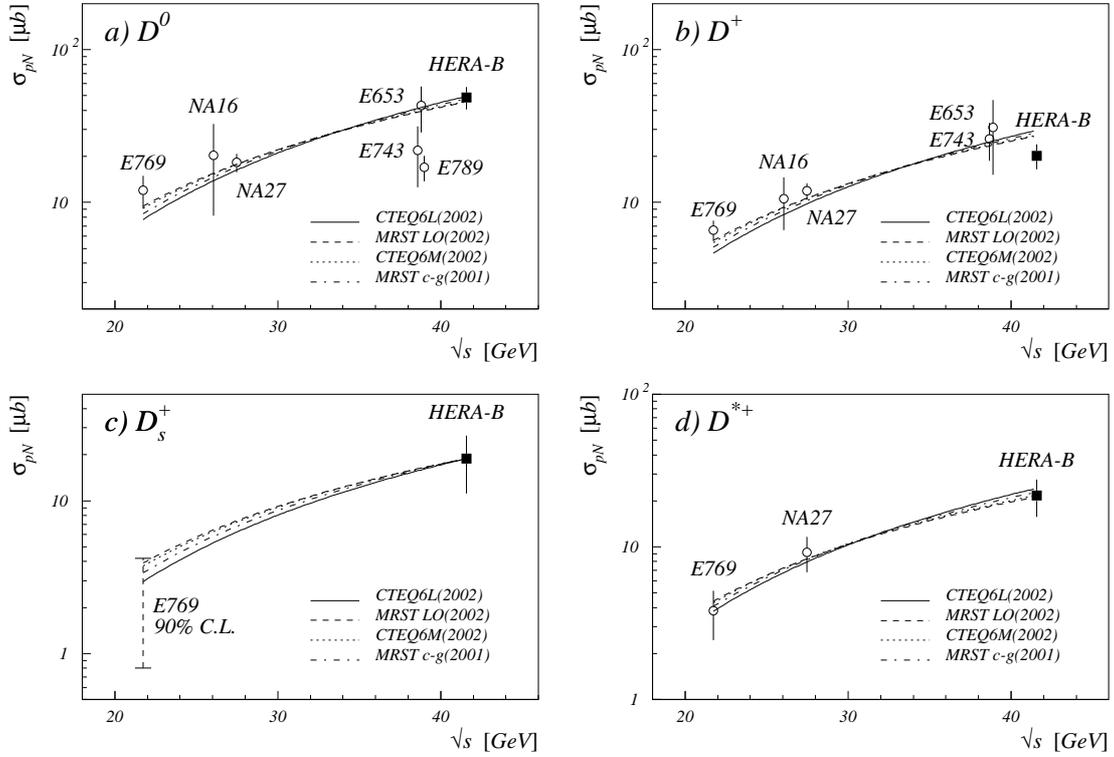}}
  \caption{Comparison of the present results with previous 
measurements~\cite{NA16,NA27,E743,E653,E789,E769}. 
In figure a) the result of E789 is excluded from the fit. The different 
curves correspond to different assumptions on parton distribution functions~\cite{lourenco}.}
  \label{xsec-final.eps}
\end{figure*}

Figure \ref{xsec-final.eps} shows the comparisons of these measured cross sections with other 
experimental studies. To compare the data points at different
center-of-mass energies, the overall normalization of theoretical 
predictions given in \cite{lourenco} is determined separately 
for each of the $D$ mesons by fits to the data which also include
the results of the present measurements \footnote{The original measured values are rescaled 
in \cite{lourenco} by taking into account the updated $D$ branching fractions.
 The result of $\sigma(D^0)$ by E789 is excluded from the fit.}.
Our results are consistent with previous studies and represent an improvement 
at high energies.
 Note that our   $\sigma(D^+)$ is somewhat lower than expected, leading to a lower ratio
   $\sigma(D^+)/\sigma(D^0)$ as will be discussed below.




The measured sum of the $D$ meson cross sections per nucleon 
$\sigma(D^0)+\sigma(D^+)+\sigma(D^+_s)=(87.4\pm8.2\pm12.6)$~$\mu$b is used to 
determine the charm production cross section. 
From the  fractions of 
charm hadrons produced in the hadro\-nization of $c$ quarks as measured in $e^+e^-$ collisions,
the production of $D^0$,  $D^+$ and $D_s$ mesons is found to account for 
$f_D=0.891\pm0.041$ of the charm cross section \cite{PDG200}, where the uncertainty
 in $f_D$ 
is obtained from errors in individual fractions neglecting possible correlations.  
 Assuming the same fraction, the present study derives the charm cross section per nucleon,
$\sigma(c\bar c)=(\sigma(D^0)+\sigma(D^+)+\sigma(D^+_s))/(2 f_D)$,
where the factor 2 accounts for the charge-conjugated states which 
are included in the $D$ production cross sections.
The resulting  charm cross section per nucleon  is thus $\sigma(c\bar c)=(49.1\pm4.6\pm7.4)$~$\mu$b. 
Note that due to correlations, the systematic error
in the sum of cross sections is larger than the value which one would get by adding in 
quadrature the individual contributions.

\subsection{Differential cross sections}
\label{differential.sec}

\begin{figure}
     \centerline{\includegraphics[width=7.5cm]{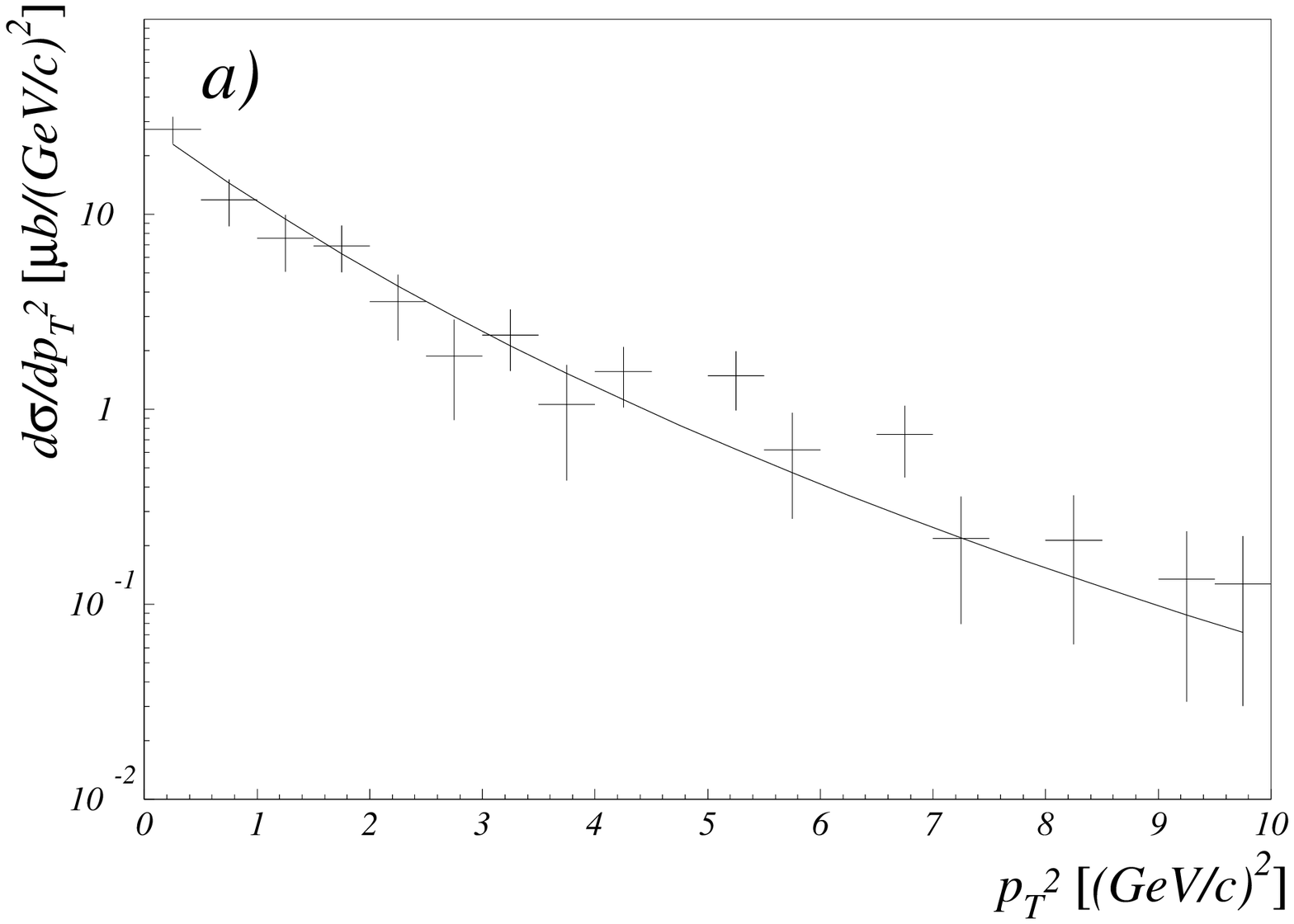}}
   \label{dsigma_dpt.eps}
     \centerline{\includegraphics[width=7.5cm]{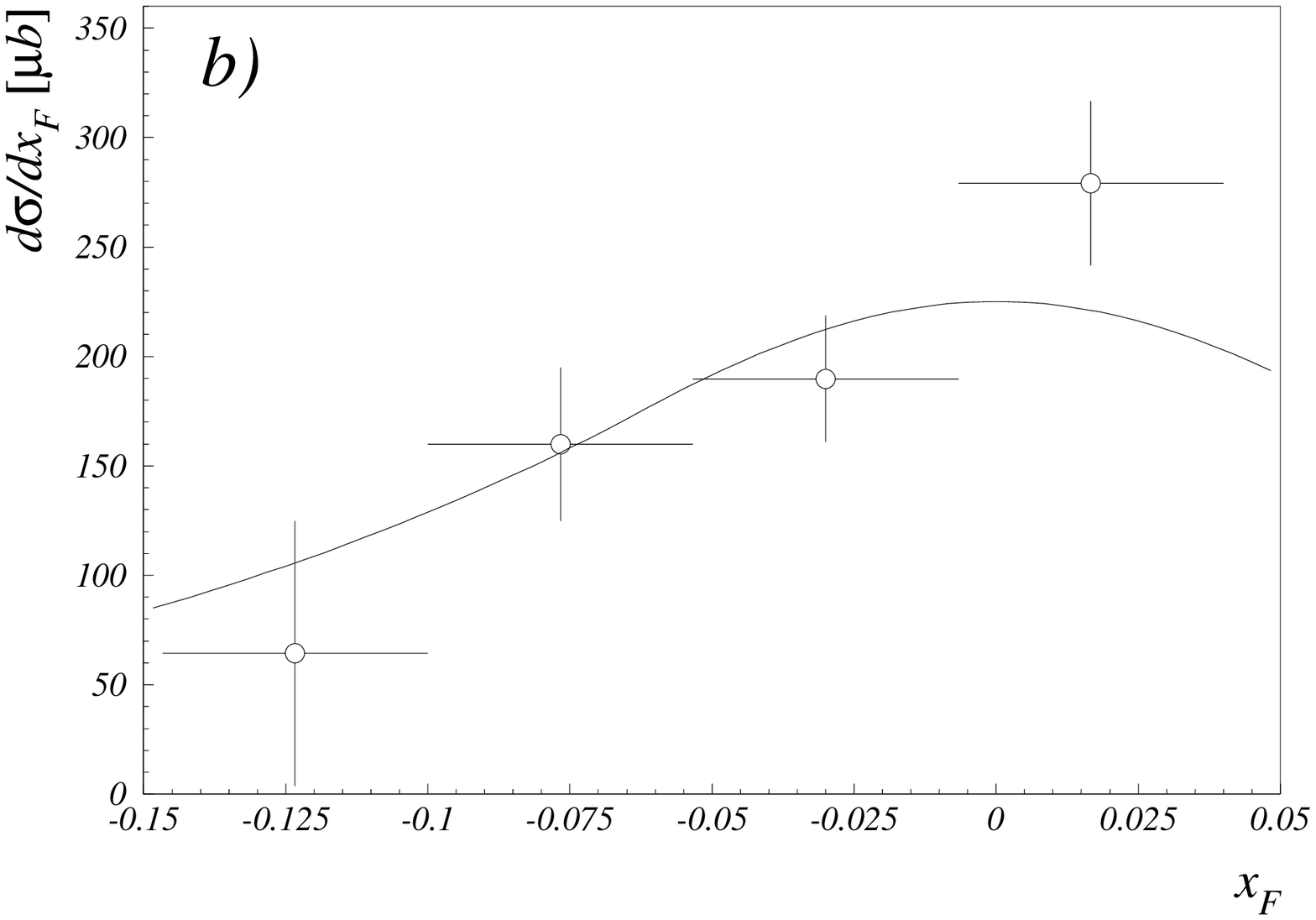}}
   \label{dsigma_dxf.eps}
   \caption{Differential visible cross sections ($-0.15<\xF<0.05$)
     for $D^0$ and $D^+$ production: (a) $d\sigma/d\pTsq$, with the fit of
     Eq.~\ref{ptform.eq}, 
     and (b) $d\sigma/d\xF$ with the fit of Eq.~\ref{xf-extended.eq}
      with a free parameter $n$ and a fixed  boundary parameter $\xb=0.062$.
   } 
   \label{dsigma_dpt_dxf.eps}
 \end{figure}

The differential cross sections  $d\sigma/d\pTsq$ and $d\sigma/d\xF$
are determined from production yields in bins of $\pTsq$ and $\xF$ 
by using Eq.~\ref{Dxsec.eq}. 
The yield in each individual bin is determined by subtracting from the 
number of events in the $D$ meson signal window the number of background 
events as estimated from the sidebands.
The resulting  differential cross sections for the production of  $D^0$ or $D^+$
mesons are shown in Fig.~\ref{dsigma_dpt_dxf.eps}.

The parameters of the measured differential cross sections are determined in the following 
way. Because of the low statistics in individual  bins of $\pTsq$ and $\xF$, 
we do not directly fit the background-subtracted distributions but instead
do simultaneous binned likelihood fits of the
 $\pTsq$ and $\xF$
 distributions of events in the mass signal and  sideband windows.
 This correctly accounts for the Poisson errors.

For the transverse momentum distribution, several pa\-ra\-meterisations 
can been found in the literature~\cite{frixi,E791,E769-1}. In the present
analysis, the parameterisation given in Eq.~\ref{ptform.eq} is used, 
since  our previous studies of $J/\psi$, $K^*$ and $\phi$
production indicate that it fits the data  well over a large range of  $\pTsq$.
The values of parameters 
$\lvec \pT \rvec$ and $\beta$ are extracted from the fit.
The distribution of background events $N_{\rm bgr}(\pTsq)$ are assumed to be the same
in shape and normalization for the signal window and sidebands. Several
parameterisations
of the distributions of events in sidebands  fit
well. We use the parameterization with two free parameters  which gives the smallest
$\chi^2/$ndf: 
$Ce^{B\pTsq }$ for $D^0$ and $D^+$. The same parameterisation multiplied 
by the efficiency in $\pTsq$, $Ce^{B\pTsq }\epsilon(\pTsq )$, is used for $D^{*+}$.

The $D^0$, $D^+$ and $D^{*+}$ data samples are fit simultaneously, where in the $D^{*+}$ 
case only the subsample with $D^0$ daughters not 
  common to the $D^0$ sample is included.
The resulting fit parameters are $\lvec \pT \rvec = (1.04\pm0.04)$~GeV/$c$ and 
$\beta = 7.0\pm4.3$, with $\chi^2/$ndf=0.86. 
The measured  $\lvec \pT \rvec$ is significantly larger (by 3.3$\sigma$) than 
the value extracted from the Pythia Monte Carlo samples (0.90~GeV/$c$),
while $\beta$ is, within one standard deviation, equal to the 
value extracted from the  simulation ($\beta=4.80$).  
 Fitting with the fixed value of $\beta=6$, i.e., the value  
used for our study of $J/\psi$ production \cite{jpsi} yields
the same $\lvec \pT \rvec$ value, and the same $\chi^2/$ndf.
%
%
%
 
The $\xF$ distribution is usually parameterised with a power-law function:
\begin{eqnarray}
  \frac{d\sigma}{d\xF} \propto (1-|\xF|)^n.
  \label{xf-usual.eq}
\end{eqnarray}
This function differs in the central region from 
the predictions of the next-to-leading order QCD calculations~\cite{nason89}. The 
 measurements made by E791 \cite{E791} in 500~GeV $\pi$A collisions 
using a high statistics sample of 80k reconstructed 
$D^0$ mesons also show a similar discrepancy. They obtain an improved fit with the function
given in Eq.~\ref{xf-extended.eq},   which
uses a power-law function in the tail region and a Gaussian in the central 
region\footnote{To account for the asymmetry in $\pi$-A collisions, E791 
used an additional offset parameter}.
 
In our analysis of the measured differential cross section $d\sigma/d\xF$ 
(Fig.~\ref{dsigma_dpt_dxf.eps}(b)), 
the boundary parameter
$\xb$ was fixed to the value $\xb=0.062$ as measured by E791 \cite{E791}, because our range of
$-0.15<\xF<0.05$ is too small to determine this parameter.
The exponent $n$, is extracted by simultaneously fitting the $D^0$ and $D^+$ data samples with a
likelihood fit to the events in the signal windows and sidebands.
While the fitted value of the exponent $n=7.5\pm3.2$ agrees with the results of
 E653~\cite{E653} and E743 \cite{E743}, the statistical error is larger in our study. 

\subsection{Ratios of cross sections}
\label{ratios.sec}

\begin{figure}[t]
  \centerline{\includegraphics[width=8cm]{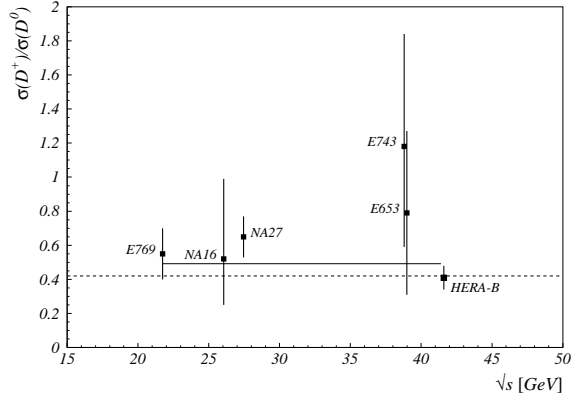}}
  \caption{Cross section ratio $R=\sigma(D^+)/\sigma(D^0)$, comparison with previous 
experiments. The dotted line at $R=0.42$ shows the prediction of the 
isospin model with $P_{\rm V}=0.6$ \cite{lourenco}; the solid line is a fit to the data points.}
  \label{xsec-ratio.eps}
\end{figure}

The measured cross section ratios are summarized in Table \ref{xsec-ratio.tab}.
The systematic errors come  mainly from selection criteria, 
event fitting, branching fraction uncertainties 
and re-weighting, while the luminosity error cancels. 
The value for the ratio $\sigma(D^+)/\sigma(D^0)=0.41\pm0.06\pm0.04$ is the most accurate
measurement of this ratio in $pA$ reactions. 
It is in good agreement with the combined results from hadroproduction~\cite{lourenco} as
 well as 
from $e^+e^-$ experiments~\cite{PDG}. The ratio also agrees 
with a simple prediction
based on isospin symmetry and the measured ratio of vector to scalar meson production 
cross sections \cite{lourenco}. 
A comparison with results from other 
experimental studies is presented in Fig.~\ref{xsec-ratio.eps}.

 The ratio  $\sigma(D^{*+})/\sigma(D^0)=0.44\pm0.11\pm0.05$ is also the most precise 
measurement of this ratio  in $pA$ reactions and is in good agreement with the 
results of NA27 and E769~\cite{NA27,E769}. 
The vector to scalar meson production ratio $P_{\rm V}$ 
can be calculated in several ways (see \cite{lourenco} 
and references therein) if isospin invariance is assumed.
From the ratio $R_1=\sigma(D^{+})/\sigma(D^0)$  one obtains 
$P_{\rm V}=(1-R_1)/((1+R_1){\rm Br}(D^{*+}\to D^0 \pi^+))=0.61\pm0.09\pm0.06$. As a
cross-check, we determine the same ratio from  $R_2=\sigma(D^{*+})/\sigma(D^+)$, and obtain  
the value $P_{\rm V}=R_2/(1+{\rm Br}(D^{*+}\to D^0 \pi^+)\cdot R_2)=0.62\pm0.09\pm0.05$.
 The results are in 
good agreement with the world average value $P_{\rm V}=0.59\pm0.01$~\cite{lourenco}.

Our result for the ratio $\sigma(D_s^{+})/(\sigma(D^0)+ \sigma(D^+))=0.27\pm0.09\pm0.05$ 
is the first measurement of this quantity in $pA$ reactions. 
For comparison, 
  the world average value  of measurements in $e^+e^-$ collisions is 0.10$\pm$0.02~\cite{PDG},
and $0.112^{+0.024}_{-0.020}$ in deep inelastic scattering at HERA~\cite{ZEUS}.
%
\begin{table}
  \caption{Ratios of cross sections.  The first error is statistical and 
the second systematic.}
  \begin{center}
    \begin{tabular}{@{}lcc}
      \br
      $D^+/D^0$         & 0.41$\pm$0.06$\pm$0.04 \\
      $D^{*+}/D^0$      & 0.44$\pm$0.11$\pm$0.05 \\
      $D^+_s/(D^++D^0)$ & 0.27$\pm$0.09$\pm$0.05 \\
      $D^{*+}/D^+$      & 1.07$\pm$0.26$\pm$0.14 \\
      \bbr
    \end{tabular}
  \end{center}
  \label{xsec-ratio.tab}
\end{table}

\subsection{Leading to non-leading particle asymmetries}
\label{leading.sec}

\begin{table}
  \caption{Leading to non-leading particle asymmetries $\mathcal{A}$
   in the visible range, $-0.15<\xF<0.05$. The first error is statistical and 
the second systematic. }
   \begin{center}
    \begin{tabular}{@{}ccc}
      \br
      $D^0$ & $D^+$ & $D^{*+}$  \\
      \mr
             0.10$\pm$0.09$\pm$0.05
      &      0.14$\pm$0.11$\pm$0.06
      &      0.23$\pm$0.17$\pm$0.06 \\
      \bbr
    \end{tabular}
   \end{center}
  \label{lead-nonlead.tab}
\end{table}
A leading particle is defined as one which has a light quark in common 
with 
the beam particle, in our case: anti-$D$ mesons
$\overline{D}^0$, $D^-$ and $D^{*-}$.
The leading to non-leading particle asymmetry is defined as
$\mathcal{A} \equiv (\sigma_{\mbox{\scriptsize LP}}
       -\sigma_{\mbox{\scriptsize non LP}})/(\sigma_{\mbox{\scriptsize LP}}
       +\sigma_{\mbox{\scriptsize non LP}})$.
The measured values of the asymmetry in the visible range, $-0.15<\xF<0.05$,
 are given in Table \ref{lead-nonlead.tab}
for the $D^0$, $D^+$ and $D^{*+}$ mesons. 
The systematic uncertainty is dominated by the uncertainty of selection criteria and the kinematical 
re-weighting with respect to possible differences between leading and non-leading particles.
Our measurements can be compared to the existing measurements of 
this asymmetry in different  $\xF$ intervals:
by E769  (0.06$\pm$0.13,  0.18$\pm$0.11 and 0.36$\pm$0.26 
for $D^0$, $D^+$ and $D^{*+}$ mesons with $\xF>0.0$ \cite{E769}), and 
 E789  (0.02$\pm$0.06 for $D^0$  mesons with $0.0<\xF<0.08$ \cite{E789}).


\subsection{Atomic mass number dependence}
\label{adep.sec}

From the measured $D$ meson production cross sections on three  different target materials 
(Table \ref{xsec-mat.tab}), the exponent $\alpha$ of Eq.~\ref{A-dep.eq} can be 
determined. The results of simultaneous maximum likelihood fits to the invariant mass distributions of individual material data samples for each $D$ meson
are summarized in Table \ref{Adep-alpha.tab}.  
The systematic error has two contributions: the uncertainty in the luminosity per
wire material (about 2.2\%), and Monte Carlo statistics (less than 1\%).
The observed value,  $\alpha=0.99\pm0.04\pm0.03$, 
 is compatible with a linear dependence of the cross sections on 
atomic mass number ($\alpha=1$). 
Note that for the weighted average of $\alpha$ over
all four samples, only those $D^{*+}$ events  were considered  for which
the $D^0$ daughter particles were not reconstructed in the $D^0$ sample. 
Our result is in agreement
with the result of E789 \cite{E789}, $\alpha=1.02\pm0.03\pm0.02$.

\begin{table}
  \caption{Atomic mass number dependence parameter $\alpha$ 
     and its weighted average for the four $D$ mesons.
    For the weighted average the value in parenthesis of the fifth 
   row were used, corresponding to 
   the subsample of $D^{*+}$ with $D^0$ daughter not 
    common with the $D^0$ sample. 
  }
  \begin{center}
    \begin{tabular}{@{}lll}
      \br
      Particle & $\phantom{xxxxxxxx} \alpha$ \\
      \mr
      $D^0$     & \phantom{(}0.969$\pm$0.057$\pm$0.026 \\
      $D^+$     & \phantom{(}1.051$\pm$0.082$\pm$0.028 \\
      $D^+_s$   & \phantom{(}1.190$\pm$0.402$\pm$0.046 \\
      $D^{*+}$  & \phantom{(}0.832$\pm$0.138$\pm$0.022 \\
                &(0.847$\pm$0.185$\pm$0.022) \\
      \mr
      Average   & \phantom{(}0.994$\pm$0.044$\pm$0.025 \\
      \bbr
    \end{tabular}
  \end{center}
  \label{Adep-alpha.tab}
\end{table}

\section{Summary}

With the HERA-B detector we have measured the total and single differential
cross sections $\sigma$, $d\sigma/d\pTsq$ and $d\sigma/d\xF$, 
the atomic mass number dependence 
of the cross sections, and the leading to non-leading particle asymmetries for 
the production of $D^0, D^+, D_s^+$ and $D^{*+}$ mesons in $pA$ collisions at 
the proton energy of 920~GeV. 

Extrapolating to the full phase space,   the total cross sections per nucleon (in $\mu$b) 
are:
$48.7\pm4.7\pm6.6$,
$20.2\pm2.2\pm3.0$,
$18.5\pm6.4\pm4.1$ and 
$21.6\pm4.7\pm3.6$  for the $D^0$, $D^+$, $D_s^+$ and $D^{*+}$,
respectively. In the 
 range $-0.15<\xF<0.05$ the measured cross sections are:
 $26.8\pm2.6\pm2.7$, 
 $11.1\pm1.2\pm1.3$, 
 $10.2\pm3.5\pm2.0$ and 
 $11.9\pm2.6\pm1.7$ for the $D^0$, $D^+$, $D_s^+$ and $D^{*+}$, respectively.
 The cross section per nucleon for $c\bar c$ production is
 $\sigma(c\bar c)=(49.1\pm4.6\pm7.4)$~$\mu$b.

We have measured the cross section ratios 
$\sigma(D^+)/\sigma(D^0)$ $=0.41\pm0.06\pm0.04$ and 
$\sigma(D^{*+})/\sigma(D^0)=0.44\pm0.11\pm0.05$, as well as the
vector to scalar meson production ratio,  $P_{\rm V}=0.61\pm0.09\pm0.06$.
Our result for the ratio 
$\sigma(D_s^{+})/(\sigma(D^0)+ \sigma(D^+))=0.27\pm0.09\pm0.05$ is the first 
measurement of this quantity in 
$pA$ reactions.

From the measured atomic mass number  dependence of the production
cross section, the parameter $\alpha=0.99\pm0.04\pm0.03$ is
extracted. This value is in 
agreement with the assumption of a linear dependence of cross sections, 
$\alpha$=1.
The measured leading to non-leading particle asymmetries in the $\xF$ range $-0.15<\xF<0.05$ 
are consistent with existing measurements for different $\xF$ regions.

The results of our studies are in good agreement with previous 
measurements of open charm production in  $pA$ interactions
and provide, in the majority of cases, an improvement in accuracy.

\section*{Acknowledgments}

We express our gratitude to the DESY laboratory for the strong support in setting up and
running the HERA-B experiment. We are also indebted to the DESY accelerator group for
their continuous efforts to provide good and stable beam conditions. The HERA-B experiment
would not have been possible without the enormous effort and commitment of our technical and
administrative staff. It is a pleasure to thank all of them.

\end{document}